\documentstyle[epsfig,12pt]{article}
\begin{document}
%\large

\begin{center}
{\Large \bf
Wavelet analysis of angular distributions of secondary particles in
high energy nucleus-nucleus interactions}
\end{center}
\begin{center}
{\bf
Irregularity of particle pseudorapidity distributions}
\end{center}

\begin{center}
V.V.Uzhinskii, V.Sh.Navotny\footnote{FTI, Tashkent, R. Uzbekistan},
G.A.Ososkov, A.Polanski, M.M.Chernyavski\footnote{FIAN, Moscow, Russia}
\end{center}

\begin{center}
Joint Institute for Nuclear Research, \\
Laboratory of Information Technologies
\end{center}

\begin{center}
\begin{minipage}{13cm}
Experimental data on sulphur and oxygen nuclei interactions
with photoemulsion nuclei at the energies of 200 and 60 GeV/nucleon are
analyzed with the help of a continuous wavelet transform.
Irregularities in pseudorapidity distributions of narrow groups of the
secondary shower particles in the mentioned interactions are observed at
application of the second order derivative of Gaussian as a wavelet.
The irregularities can be interpreted as an existence of the preference
emission angles of groups of particles. Such an effect is expected at
emission of Cherenkov gluons in nucleus-nucleus collisions. Some of
the positions of the observed peculiarities on the pseudorapidity axis
coincide with those found by I.M.Dremin et al.  (I.M.Dremin et al. Phys.
Lett., 2001, v. B499, p.  97).  )
\end{minipage}
\end{center}

In the last decade a new type of mathematical analysis of data,
the so-called wavelet analysis has become very popular in various
branches of science and engineering \cite{W1}--\cite{W4}. Mainly, it is
applied for analysis of time series coming from geophysics,
meteorology, astrophysics, and so on (applications in aviation, medicine
and biology see in \cite{Dremin_r}).

The wavelet decomposition or wavelet transform of a function $f(x)$ is
its decomposition on an orthogonal functional family of special form
\cite{Daubechi};
\begin{equation}
W_{\Psi}(a,b)f=\frac{1}{\sqrt{C_{\Psi}}} \int ^\infty _{-\infty}
f(x) \Psi _{a,b}(x) dx,
\label{eq1}
\end{equation}
$$
\Psi _{a,b}(x) \equiv a^{-1/2} \Psi\left( \frac{x-b}{a} \right),
$$
where $\Psi$ is called a wavelet, $b$ -- a translation parameter, $a$ --
a dilation parameter or a scale, and $C_{\Psi}$ -- a normalizing
constant
$$
C_\Psi =2\pi \int _{-\infty}^{+\infty}|\tilde{\Psi }(\omega )|^2 |\omega |^{-1}d\omega ,
$$
where $\tilde{\Psi }(\omega )$ -- Fourier transform of $\Psi (x)$.

The derivatives of the Gaussian function are often used as wavelets
$$
\Psi (x)\equiv g_n(x)=
(-1)^{n+1}\frac{d^n}{dx^n}\mbox{e}^{-x^2/2}, ~~~ n > 0,~~~ C_{g_n}=2
\pi (n-1)!.
$$
The first two wavelets are well known:
$$
g_1(x)=-x\mbox{e}^{-x^2/2}, ~~~ g_2(x)=(1-x^2)\mbox{e}^{-x^2/2}.
$$
The second one is called Mexican Hat wavelet (MHAT).

As seen, the wavelet transform puts in correspondence to the function
of one variable, $f(x)$, the function of two variables, $W_{\Psi}(a,b)$.
Until recently, a presentation and an analysis of a function of
two variables was quite a difficult job, and only modern computers with
their 3D-graphics allowed one to implement completely the method of
wavelet analysis.

There is a discrete analogy of the continuous wavelet transform (see
\cite{Daubechi}). It was used in elementary particle physics for the
study of events of cosmic rays interactions with materials -- at the
analysis of particle pseudorapidity distributions \cite{Suzuki}. The
wavelet coefficients ($W_{\Psi}(a,b)$) of energy distributions
predicted by different models of multi-particle production were studied
in \cite{DHuang}. Possibilities of the wavelet analysis in searching
for manifestation of the disoriented chiral condensate in the
pseudorapidity distributions of neutral particles fraction were
considered in \cite{ZHuang}.

In papers \cite{Dremin} the wavelet transform was used for pattern
recognition in $Pb+Pb$-interactions at the energy of 158 GeV/nucleon.
Structures were found in the angular distributions of secondary
particles that can be interpreted as irradiation of Cherenkov gluons.
The last publications in this research are devoted to
experimental search for the disoriented chiral condensate in
nucleus-nucleus interactions \cite{Kopytin}.

Most of the mentioned papers suffer from the apparent lack of
quantitative results which caused either by uniqueness of nature
phenomenon or low statistics of analyzed data. It is connected partly
with specific properties of the wavelet analysis itself (see
\cite{Torrence}). A regular method of wavelet analysis application in
particle physics is needed. Our paper presents experience of using
the wavelet transform for analysis of more than 2000 interactions of
nuclei with nuclei at high energies.

Experimental data were obtained at horizontal irradiation of NIKFI BR-2
nuclear photoemulsion by sulfur and oxygen nucleus with the energies
of 200 and 60 GeV per nucleon at the CERN SPS.
The sensitivity of emulsion was about 30 grains per unit length of 100
$\mu m$ for single charged particles with minimal ionization.

Primary interactions were found by along-the track double scanning:
fast in the forward direction and slow in the backward direction. Fast
scanning was made with a velocity excluding any discrimination of
events in the number of heavily ionizing tracks, slow scanning was
carried out to find events, if any, with little changed and unbiased
projectile nucleus. Upon rejecting events of electromagnetic
dissociation and purely elastic scattering in the total sample, 884
events of $S+Em$ and 504 events of $O+Em$ at the energy of 200
GeV/nucleon, and 884 $O+Em$ interactions at the energy of 60
GeV/nucleon were selected for a further analysis.  In each event the
polar angles $\theta$ and azimuthal angles $\varphi$ were measured.

Shower particles, or the so-called $s$-particles -- single charged
particles with a velocity $\beta = v/c \geq 0.7$, are considered in
this study.  The $s$-particles consist mainly of produced particles and
single charged nuclear fragments. A special separation of single
charged fragments was not done.

According to Ref. \cite{Dremin}, a distribution of the secondary
particles on the pseudorapidity $\eta =-\ln (\tan (\theta /2))$ in an
event was presented as
\begin{equation}
f(\eta )=\frac{dn}{d\eta}=\frac{1}{N}\sum
_{i=1}^N \delta (\eta - \eta _i), \label{eq2}
\end{equation}
where $N$ is multiplicity of $s$-particles in the event, and $\eta _i$
is pseudorapidity of i-th particle.

A wavelet transform of the function (\ref{eq2}) gives
\begin{equation}
W_{\Psi}(a,b)=\frac{1}{N}\sum _{i=1}^N a^{-1/2} \Psi\left(
\frac{x-b}{a} \right).
\end{equation}
So, the function of two variables is brought into correspondence with
each particle. Wavelet spectra of the event ($W_{\Psi}(a,b)$) is a sum
of such functions.

As an example, Fig. 1 presents wavelet spectra of event with 6
particles having pseudorapidities $\eta _1=1$, $\eta_2=2.75$,
$\eta _3=3.25$, $\eta _4=5$, $\eta _5=6$, $\eta _6=7$. The values of
translation parameter $b$ are put on the X-axis, dilation parameter,
$a$ -- on Y-axis. The values of the wavelet coefficients are depicted
in a gray-level scale: the high values of the coefficients are of
light shade while the lower ones are darker. As seen, in $g_2$ wavelet
spectra at the scale lower than 0.3 all particles are distinguished. At
the scale larger than 0.5, the particles 2 and 3 can not be resolved.
At $a > 1$ particles 4, 5, 6 can not be resolved. At $a >2$ one
could expect a fusion of particles 1, 2, and 3. However, this does
not take place due to small yield of particle 1 in the wavelet
spectra at a large scale. So, the wavelet transform of $g_2$ allows one
to study the particle clusterization.
\begin{figure}[cbth]
\begin{center}
\psfig{file=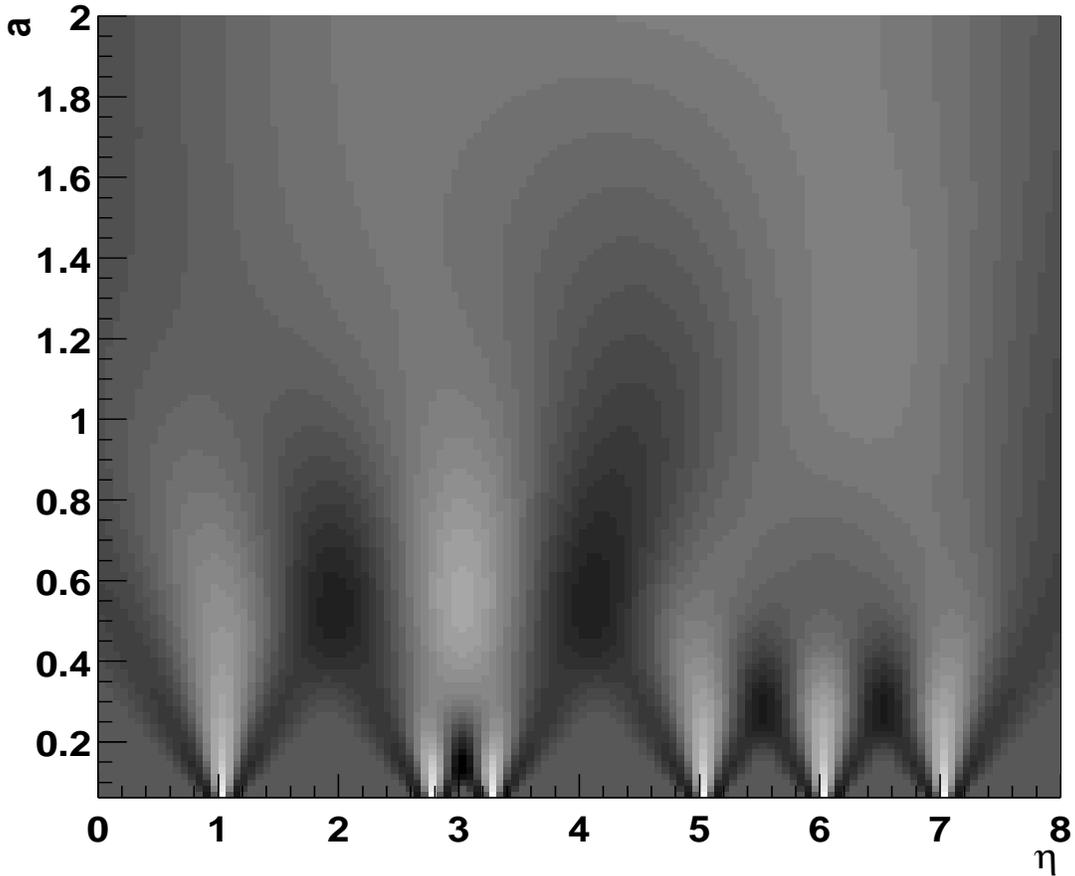,width=160mm,height=130mm,angle=0}
%\vspace{130mm}
\caption{Wavelet spectra, energy densities, and scalogramms}
\end{center}
\end{figure}
%\vspace{-0.4cm}

Let us note that the positions of the local maximums of $W_{\Psi}(a,b)$
($a \simeq 0.5$ and $b=3$, $a \simeq 1$ and $b=6$) on the b-axis are
connected with positions of the centers of the groups of particles
2, 3 and 4, 5, 6. The corresponding scales reflect the width of the
groups on the pseudorapidity axis. Positions of the local minimums give
the centers of the pseudorapidity splits between the groups.

There are a lot of wavelets. Different representations of the 6
particles of the test example are shown in Fig. 1. There are also
energy densities ($W_{\Psi}(a,b)^2$) and scalogramms $ E_W(a) = \int
W_{\Psi}(a,b)^2 db, $ which are often used in practice.

The $g_1$ wavelet (see (2)) has a minimum at a negative value of its
argument and maximum -- at positive value. Positions of these extreme
points at different scales can be seen quite well in the energy density
spectra.  The $g_2$ wavelet has 3 extreme points. All of them are
presented in the energy spectra (see the region of the 1 particle).
Scalogramms give generalized imagination of the extreme points.

As seen in the figure, different wavelets give different presentations
of particles. The more "usual" ones are obtained at the
application of even wavelets. Perhaps, odd wavelets can be
useful at automatic data processing.

At the first glance, the $g_1$ wavelet allows one to locate the regions
of unhomogeneouslity in the particle distribution. The positions of the
local maxima of the energy spectra on the b-axis ($b=2.5$ and
$b=3.5$) mark the group of the particles 2 and 3, and the positions on
the a-axis show the group size. Though, the following local maxima
are disposed at $a=4$ and $b \sim 0$, $b \sim 9$. There is no selection
of the group of the particles 4, 5, 6. At the same time, there are
irregularities (extreme points and inflection points) in the
corresponding scalogramm which are probably connected with
characteristics of the particle group.
                                             7
There are extreme points in the scalogramm obtained with help of the
$g_2$ wavelet at $b\simeq 0.5$ and 1.2 associated with the group
characteristics. So, we believe that scalogramms can be used for fast
search of the particle group.

The $g_4$ wavelet has analogous properties. Though, the
characteristics of the corresponding scalogramms are not so strongly
correlated with the particle characteristics. Thus, below we will use
mainly the $g_2$ wavelet.

We started our analysis with study of the particle distribution on the
pseudorapidities in all interactions. The histogrammed distribution for
$S+Em$ interactions at the energy of 200 GeV/nucleon is presented in
Fig. 2. There are also $g_2$ and $g_4$ wavelet spectra at different
scales. It should be noted that the wavelet transform was applied to
raw experimental data, not to the histogrammed distribution. Thus,
a fine structure of the spectra can be observed at small scales. The
structure becomes more regular at large scales, and turns to the
wavelet function in the limit of large $a$. At $a \sim 0.4$ one can
see 3 maxima in the $g_2$ spectra at $\eta \sim 2,~ 4$ and 8. It
seems that the maximum at $\eta \sim 2$ is connected with the particle
production in the target fragmentation region; the maximum at $\eta
\sim 4$ -- with the particle production in the central region, and the
maximum at $\eta \sim 8$ -- with spectator fragments of the projectile
nucleus. In reality, the three maxima only reflect what can be
seen with the naked eye -- the left wing of the distribution is
different from the right one. One can clearly see a change of the slope
on the right wing at $\eta \sim 6$. At smaller scales the structure is
more rich and does not allow such simple interpretation. In order to
find selected scales in the interactions and to analyze a fine
structure of the events, we turned to a study of the energy density and
scalogramm.

The spectrum of the energy density reflects peculiarities of
$W_{\Psi}(a,b)$ function and can be used for searching the particle
group. Though, definition of maxima and minima of the function is a
quite complicated task. So, at the first stage we concentrated our
attention on the energy distribution on scales, on scalogramm
because it is a 1D-function.
\begin{figure}[cbth]
\begin{center}
\psfig{file=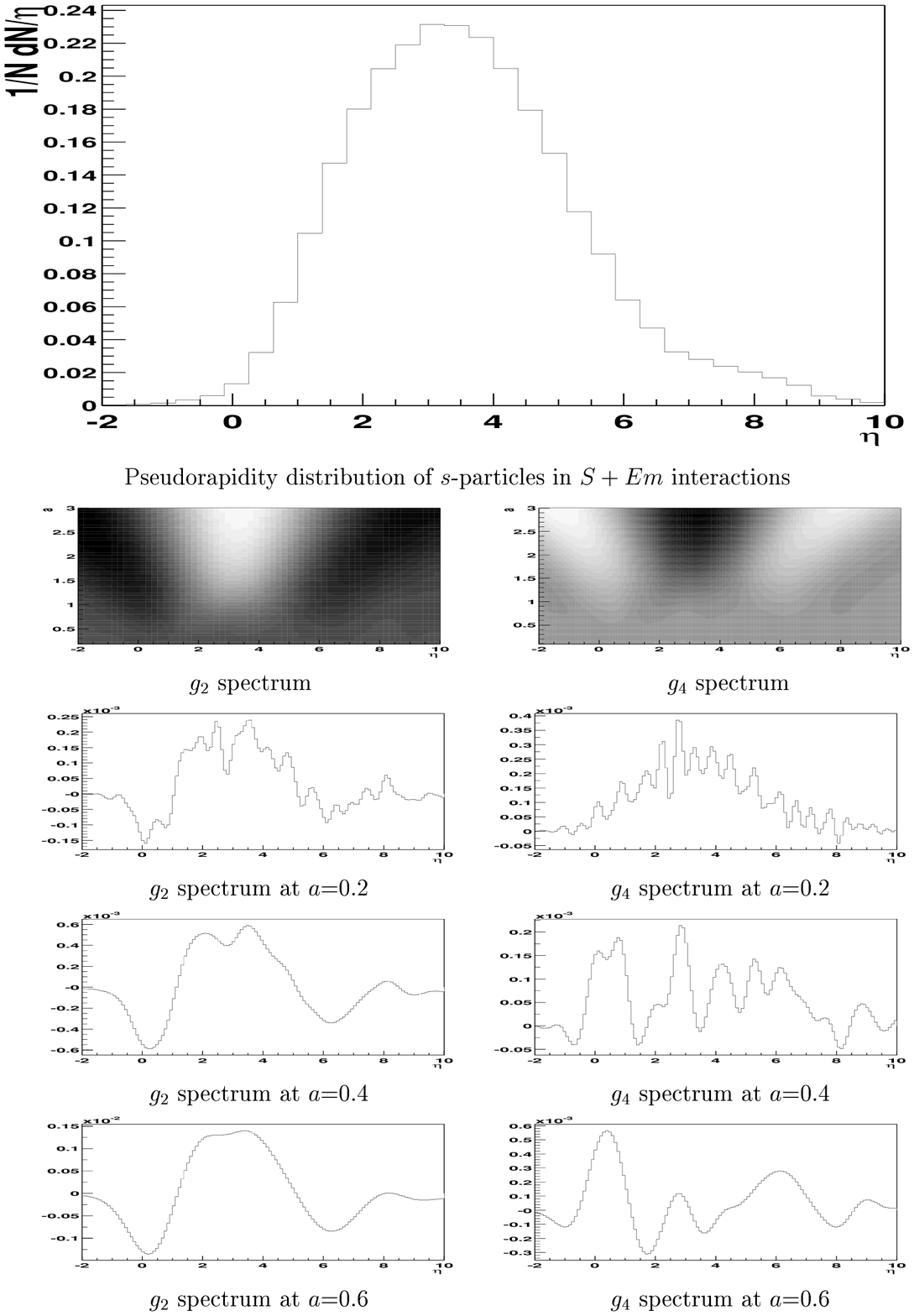,width=160mm,height=160mm,angle=0}
\caption{Wavelet analysis of $s$-particles pseudorapidity distribution
in $S+Em$ interactions at the energy 200 GeV/nucleon}
\end{center}
\end{figure}
\vspace{-0.4cm}

Scalogramms reflect characteristic features of events according to
the test example. For example, the scalogramm of the $g_2$ spectrum has
the minimum at $a \sim 1.1$ associated with an average distance between
the groups of the particles 2, 3 and 4, 5, 6, and the maximum at $a
\sim 0.5$, connected with the most compact group of particles 2, 3.
It is possible to write out an analytical expression for the scalogramm
using the $g_2$ wavelet and the distribution of the form (\ref{eq2}).
\begin{equation}
S(a)=\frac{1}{32\sqrt{\pi}a^4}\sum_{i,j=1}^N \left[ \Delta _{i,j}^4-12 a^2 \delta ^2 -12 a^4\right]
e^{-\frac{\Delta_{ij}^4}{4a^2}}
\end{equation}
$$
\Delta _{ij}=\eta_i -\eta _j
$$
It is seen that the scalogramms is a statistical averaged squared
distance between particles. Thus, we expected that an analysis of the
scalogramms allowed us to find characteristic scales.

Distributions of local minimum and maximum positions in the scalogramms
of the events of $S+Em$ interactions at the energy 200 GeV/nucleon are
given in Fig. 3. Though there are some irregularities in the
distributions, we can not say that they have a statistically guaranteed
meaning. In addition, the analysis of connection between
$a_{min}$ and $a_{max}$, and characteristics of the real events did not
show any regularity. However, we believe that the study of a larger
volume of experimental data will produce interesting results.
\begin{figure}[cbth]
\begin{center}
\psfig{file=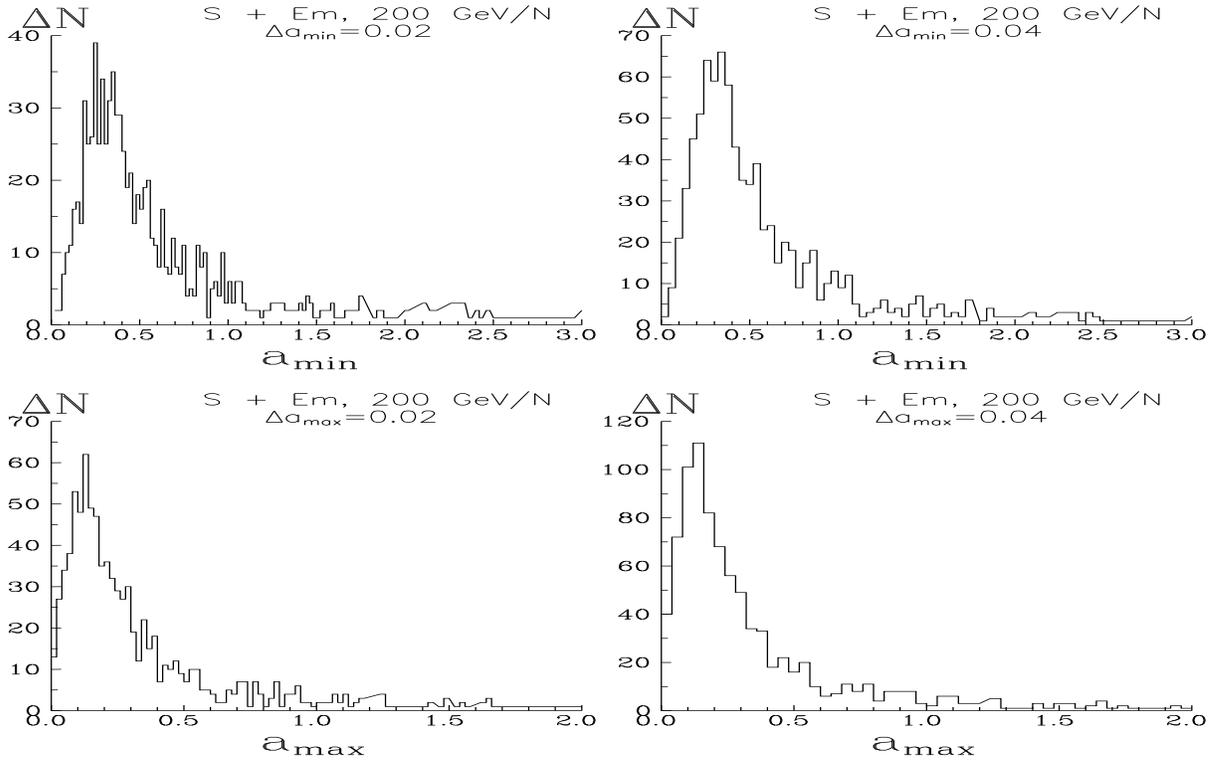,width=160mm,height=100mm,angle=0}
\caption{Distributions of the extreme points of the scalogramms of the
$S+Em$ events}
\end{center}
\end{figure}

The next step of our study was search for extremum points of
function $W_{\Psi}(a,b)$ of the real events. Distributions of points on
the dilation scale in our interactions are presented in Fig. 4. At the
first glance, the distributions have no any characteristic
peculiarities -- bumps or pits. We only can mark that the
distributions can not be described by a simple exponential function.
It is necessary to use at least 2 exponents at fitting the
distributions.  Simulation of $pp$-interactions with the help of the
HIJING program \cite{HIJING} has shown that the appearance of the
exponent with a larger slope is connected with production of jets of
particles.  So, we consider our distributions as a manifestation of jet
existence in nucleus-nucleus interactions.
\begin{figure}[cbth]
\begin{center} \psfig{file=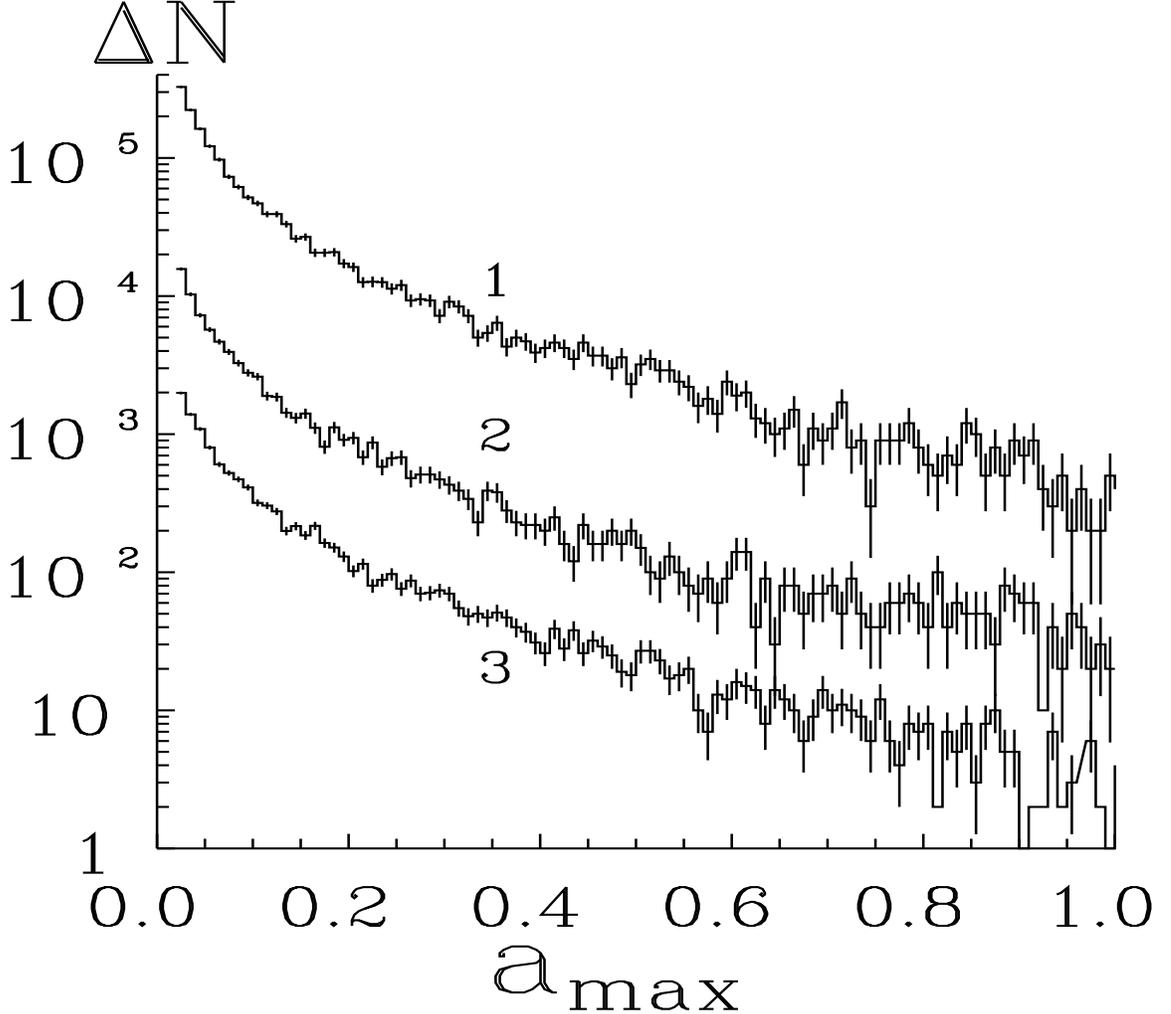,width=6in,height=5.3in,angle=0}
\caption{$a_{max}$ distributions in the $S+Em$ and $O+Em$ interactions
at the energy of 200 GeV/nucleon, and in the $O+Em$ interactions
at the energy of 60 GeV/nucleon (histograms 1, 2 and 3, respectively).
The distributions 1, 2, and 3 are multiplied by $10^2$, $10^1$ and
$10^0$, respectively.}
\end{center}
\end{figure}

Research in the distributions of local maximum of $W_{\Psi}(a,b)$ on $b$
gives more interesting results. The Fig. 4 shows the distributions of
all our interactions at $a_{max}> 0.05$ (1), $a_{max}> 0.1$ (2),
$a_{max}> 0.2$ (3), and $a_{max}> 0.3$ (4). As seen, the peculiarities
of the distributions at $\eta \sim 1.5$, $\eta \sim 2$,  $\eta \sim 3$,
$\eta \sim 3.5$, $\eta \sim 5$ are located at the same positions for
different interactions. The peculiarities, as it can be seen, are
connected with narrow groups, where $a_{max}<0.05$. We interpret them as
an existence of preference emission angles of the groups of the
particles. Let us note that the positions of some irregularities found
by us coincide with those observed early in Ref. \cite{Dremin}
at the study of $Pb+Pb$ interactions at the energy of 158 GeV/nucleon.
Unlike Ref. \cite{Dremin} where 5 mostly central events were
analyzed, we present the results for more than 2000 events.

\newpage
\begin{figure}[t]
\begin{center}
\psfig{file=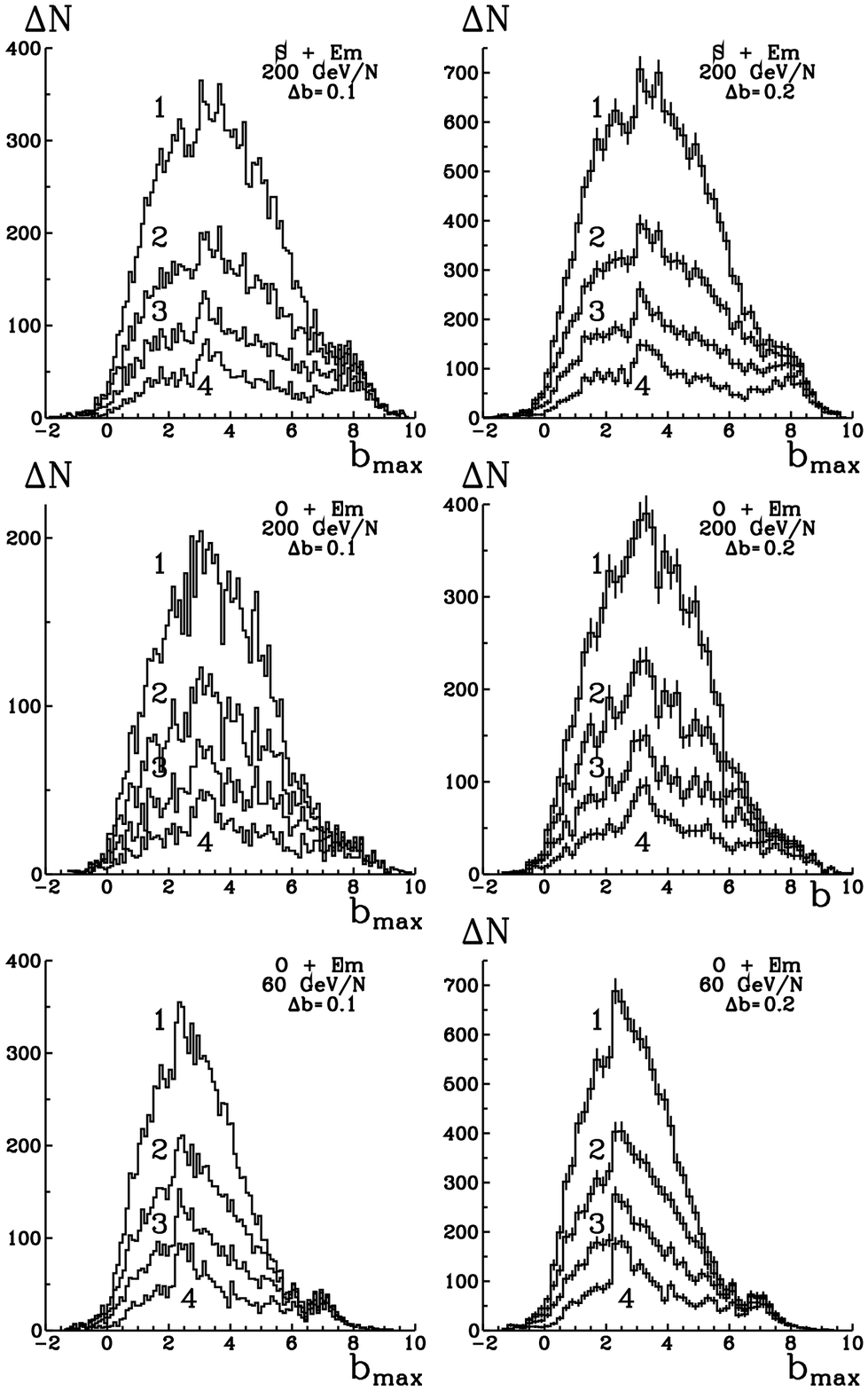,width=160mm,height=190mm,angle=0}
\caption{$b_{max}$ distributions in the $S+Em$ and $O+Em$ interactions
at the energy of 200 GeV/nucleon, and in the $O+Em$ interactions
at the energy of 60 GeV/nucleon. $\Delta b$ is step of the
histogramming. The distributions 1 -- 4 are obtained at $a_{max}\ge 0$,
$a_{max}\ge 0.05$, $a_{max}\ge 0.1$ and $a_{max}\ge 0.2$, respectively.}
\end{center}
\end{figure}
\newpage

Research on narrow groups of the particles with help of traditional
methods was the third step. Fig. 5 shows a distribution on
a pseudorapidity interval between neighboring particles in an event.
It was observed that there were pairs of the particles with close
pseudorapidities, $\Delta < 10^{-5}$! Their number increases
statistical fluctuations. The distribution of the centers of the pairs
on $\eta$ has two bumps at $\eta \sim 3$ and 4. So, we can conclude
that the peculiarities of the wavelet spectra observed by us at small
scales are connected with the narrow groups of the particles.
\begin{figure}[cbth]
\begin{center}
\psfig{file=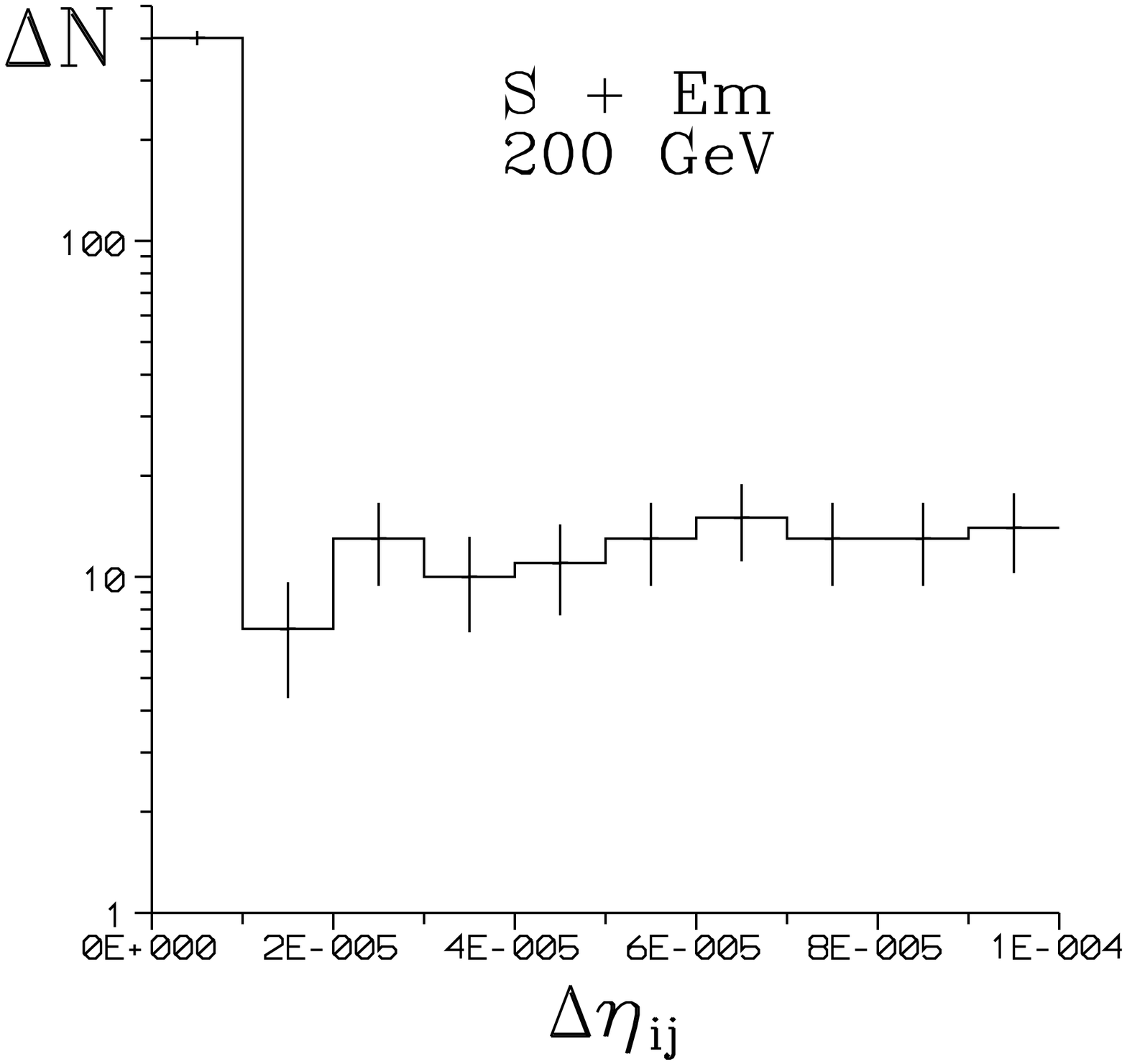,width=160mm,height=120mm,angle=0}
\caption{Distribution on the pseudorapidity interval between the
neighbouring particles in all studied interactions.}
\end{center}
\end{figure}

One can suppose that such pairs occur due to data input error when two
or three entries in the event record are corresponded to the
same particle. In this case, such ghost "particles" must have identical
$\eta$ and $\varphi$. The fraction of these pairs is lower than
20 \% among all the observed narrow pairs.  The other pairs have
different values of azimuthal angles. The distribution of the particles
of the narrow groups on $\varphi$ (see Fig. 6) has no clear
peculiarities and reflects the methodical drawback of the photoemulsion
experiments -- a poor identification of particles flying
perpendicularly to the emulsion plate (at $\varphi \sim 0^o$ and
$180^o$). The distribution of the azimuthal angle difference of the
particles of the narrow group has no irregularities either. So, these
pairs can not belong to a jet of particles. The nature of such pairs is
not clear for us.

\begin{figure}[cbth]
\begin{center}
\psfig{file=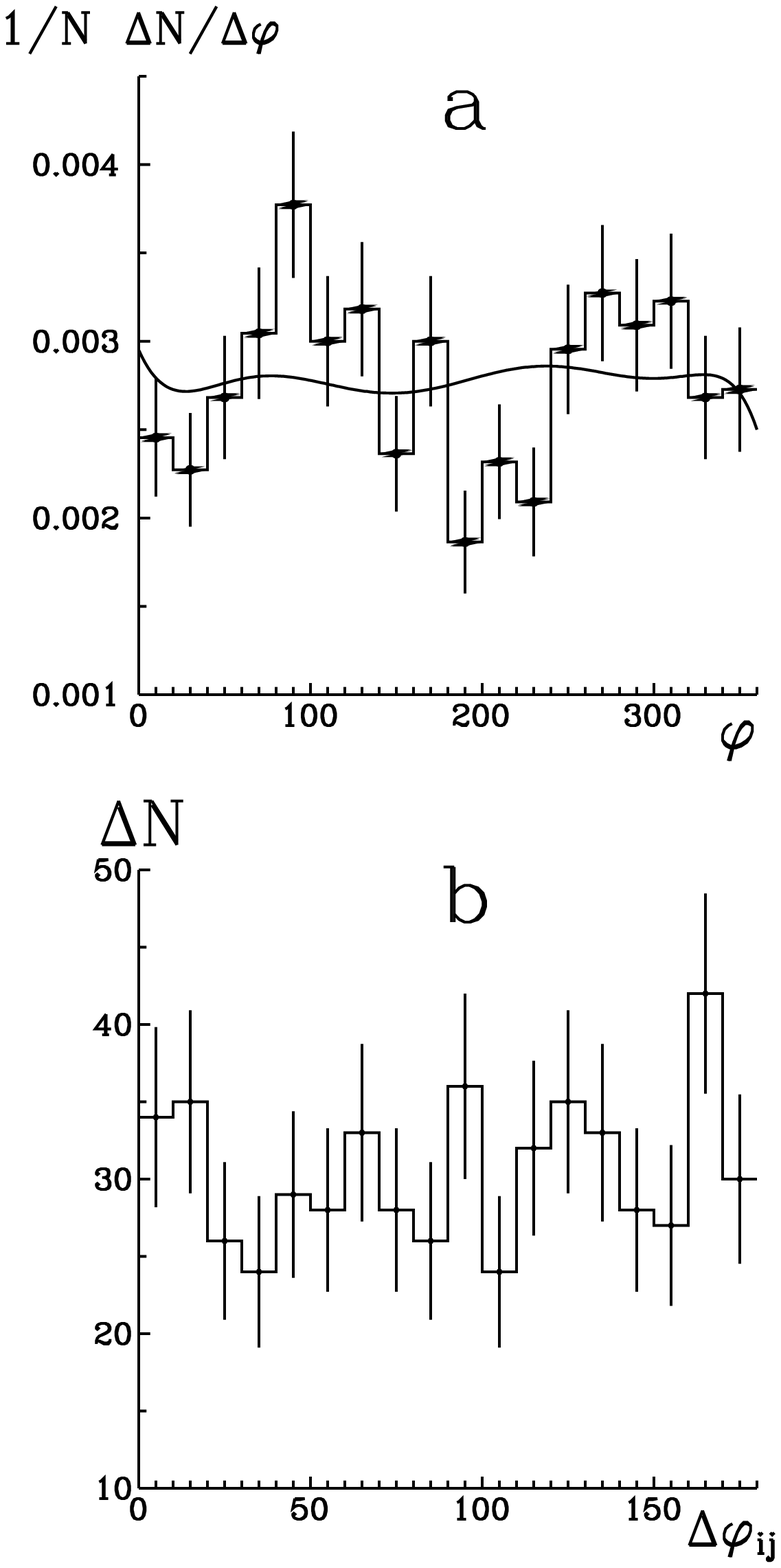,width=160mm,height=120mm,angle=0}
\caption{a) The azimuthal angle distribution of particles with close
pseudorapidities. The distribution of all particles is presented by solid line.
b) The distribution on the azimuthal angles difference of particles
with close pseudorapidities.}
\end{center}
\end{figure}

\section*{Summary}

\begin{enumerate}
\item The continuous wavelet transform has been used for the analysis
of more than 2000 events of nucleus-nucleus interactions at high
energies ($S+Em$ and $O+Em$ interactions at the energies of 200 and 60
GeV/nucleon).

\item It is shown that the maxima of $W_{\Psi}(a,b)$ obtained with
the help of the $g_2$ wavelet are associated with the groups of
particles.

\item It has been found that the distribution of the group of
particles on scales in the interactions under the study is
geterogeneous what can be caused by jet production.

\item It is observed that the
distributions of the groups on pseudorapidities have irregularities,
there are preferences of emission angles of the groups.

\item The
pairs of particles with close pseudorapidities, $\Delta < 10^{-5}$, are
found for the first time.
\end{enumerate}

The nature of the peculiarities observed by us is not clear yet.

\vspace{1cm}

The authors are grateful to the members of the EMU-01 collaboration for
their kind permission to employ the experimental data analyzed in this
paper.  One of the authors (V.V.U.) thanks RFBR (grand N 00-01-00307)
and INTAS (grand N 00-00366) for their financial support.

\end{document}